\documentclass[letterpaper,12pt,hidelinks]{article}
\usepackage{config, natbib}

\begin{document}
\title{A note on centering in subsample selection for linear regression}
\author{HaiYing Wang\\
  University of Connecticut\\
  haiying.wang@uconn.edu}
\maketitle
\begin{abstract}
  Centering is a commonly used technique in linear regression analysis. With
  centered data on both the responses and covariates, the ordinary least squares
  estimator of the slope parameter can be calculated from a model without the
  intercept. If a subsample is selected from a centered full data, the subsample
  is typically un-centered. In this case, is it still appropriate to fit a model
  without the intercept? The answer is yes, and we show that the least squares
  estimator on the slope parameter obtained from a model without the intercept
  is unbiased and it has a smaller variance covariance matrix in the Loewner
  order than that obtained from a model with the intercept. We further show that
  for noninformative weighted subsampling when a weighted least squares
  estimator is used, using the full data weighted means to relocate the
  subsample improves the estimation efficiency.
  
\end{abstract}

\section{Introduction}

In recent years, due to the challenge of rapidly increasing volumes of data, one
may have to select a small subsample from the full data so that available
computational resources at hand can fully analyze the subsample and useful
information can be drawn.   For example, the data from the second strategic
highway research program naturalistic driving study is over two million
gigabytes \citep{dingus2015naturalistic}, and existing analyses are only done on
a small proportion of it. As another example, the intelligent research in sight
registry \citep{parke2017iris}, as of 2022, has aggregated data on over four
hundred million patient visits, and it is still growing rapidly every day. An
analysis on the full data would not be possible for most practitioners due to
the super large data volume and the relatively limited computing resources.
Subsampling is also an important tool in a wide range of modern machine learning
platforms. For instance, large-scale recommender systems such as YouTube and
TikTok may receive over billions of training data points each day, creating
tremendous difficulty to effectively train online models. Subsampling is often
employed to reduce the data intensity so that online models can be updated in
time \citep{WangZhangWang2021}.

Investigations on subsampling 
have been fruitful
for linear regression. A popular technique is to use statistical leverage scores
or their variants to construct subsampling probabilities, see
\cite{Drineas:12, PingMa2015-JMLR, yang2015explicit, nie2018minimax}, and the
references therein.
\cite{WangYangStufken2019} proposed the information based optimal subdata
selection (IBOSS) method and the proposed deterministic selection algorithm has
a high estimation efficiency and a linear computational time
complexity. \cite{PronzatoWang2020} developed an online subsample selection
algorithm that achieves the optimal variance under general optimality
criteria. \cite{Yu2022} recommended using leverage scores to select subsamples
deterministically.

The slope parameter is often
the main focus and the intercept may not be of interest when fitting a linear regression model. In this scenario, a widely
used trick to simplify the calculation is to center the data so that a linear
model without the intercept can be used to calculate the slope estimator. If the intercept is needed, e.g. for prediction, it can be calculated
using the mean response and the means of the covariates together with the slope
estimator.
To be specific, consider the following linear regression model for the full data
$\Dn=(\X,\y)$ of sample size $n$,
\begin{equation}\label{eq:1}
  \y=\Z\btheta+\bve,
  =\alpha\1_n+\X\bbeta+\bve,
\end{equation}
where $\y=(y_1, ..., y_n)\tp$ is the response vector, $\Z=(\1_n\ \X)$,
$\btheta=(\alpha, \bbeta\tp)\tp$ with $\alpha$ and $\bbeta$
being the intercept and slope vector respectively,
$\bve=(\varepsilon_1, ..., \varepsilon_n)\tp$ is the model error satisfying
$\Exp(\bve)=\0$ and $\Var(\bve)=\sigma^2\I_n$, $\1_n$ is an $n\times 1$
vector of ones, and $\I_n$ is the $n\times n$ identity matrix.

To estimate $\btheta$, the ordinary least squares (OLS) estimator using the full
data $\Dn$ is
\begin{align}
  \htheta=(\halpha,\hbeta\tp)\tp=(\Z\tp\Z)^{-1}\Z\tp\y.
\end{align}
The mean response is $\bar{y}=n^{-1}\1_n\tp\y$ and the vector of
column means for $\X$ is $\bar{\x}=n^{-1}\X\tp\1_n$, so the centered data can be
written as
\begin{align}
  \X_{c}&=\X-\1_n\bar{\x}\tp=(\I_n-\J_n)\X,\\
  \y_{c}&=\y-\1_n\bar{y}=(\I_n-\J_n)\y,
\end{align}
where $\J_n=n^{-1}\1_n\1_n\tp$ and the notation $^{\otimes2}$ means
$\v^{\otimes2}=\v\v\tp$ for a vector or matrix $\v$. 
With centered data, it is well know that $\hbeta$ can be calculated as
\begin{align}
  \hbeta&=(\X_{c}\tp \X_{c})^{-1}\X_{c}\y_{c},
\end{align}
and if $\alpha$ is of interest, $\hat{\alpha}=\bar{y}-\bar{\x}\tp\hbeta$.

An interesting question raises for centering in subsampling: if the full data is
centered, do we have to center the subsample to calculate the slope estimate
if the model does not contain an intercept? We will show in this short note that it is better to not
center the subsample in this case. Since for a deterministically selected
subsample, the OLS is applied \citep[e.g.,][]{WangYangStufken2019,
  PronzatoWang2020}, while for a randomly selected subsample with nonuniform
probabilities the weighted least squares (WLS) is often fitted
\citep[e.g.,][]{yang2015explicit,ai2020optimal,zhang2021optimal}, we discuss these two types
of estimators in Sections~\ref{sec:determ-select-with} and
\ref{sec:weighted-ls}, respectively. Some numerical evaluations are provided in
Section~\ref{sec:numer-comp} and more technical details are given in the
Appendix.

\section{Deterministic selection with OLS}
\label{sec:determ-select-with}
Let ($\X^*$, $\y^*$) denote the subsample of size $r$ corresponding to the
un-centered full data $(\X,\y)$, and ($\X_{c}^*$, $\y_{c}^*$) be the
subsample corresponding to the centered full data $%
(\X_c,\y_c)$, i.e.,
($\X_{c}^*=\X^*-\1_r\bar{\x}\tp$, $\y_{c}^*=\y^*-\1_r\bar{y}$).  In this
section, we assume that the selection rule is nonrandom and it may depend on
$\X$ but it does not depend on the response $\y$.
 This type of subsampling methods includes the IBOSS that focuses on
first-order linear regression models
\citep{WangYangStufken2019}, the sequential online thinning that is designed for
online streaming data
\citep{PronzatoWang2020}, optimal design subsampling
\citep{deldossi2021optimal}, and deterministic leverage score selection for model
discrimination \citep{Yu2022}, among others. Subsampling methods in this category aim at
estimating the true parameter and they have a
higher estimation efficiency in general, but they require strong assumptions on
the correctness of the model so they should only be used when a linear
regression model fits the data well.
A subsample selected using this class of methods  follows a linear regression model
\begin{equation}\label{eq:3}
  \y^*=\Z^*\btheta+\bve^*=\alpha\1_r+\X^*\bbeta+\bve^*, 
\end{equation}
where $\Z^*=(\1_r\ \X^*)$, $\Exp(\bve^*)=\0$, $\Var(\bve^*)=\sigma^2\I_r$,
$\1_r$ is a $r\times1$ vector of ones, and $\I_r$ is the $r\times r$ identity
matrix. The OLS based on the subsample is
\begin{align}\label{eq:2}
  \ttheta=(\talpha,\tbeta\tp)\tp=({\Z^*}\tp\Z^*)^{-1}{\Z^*}\tp\y^*.
\end{align}

Clearly, $\X_{c}^*$ and $\y_{c}^*$ may not be centered, i.e., their sample means are
not zero. Can we simply use $(\X_{c}^*,\y_{c}^*)$ to fit a model without the
intercept to estimate $\bbeta$, i.e., use
\begin{align}
  \tbeta_c&=(\X_{c}^{*\tsp} \X_{c}^*)^{-1}\X_{c}^{*\tsp} \y_{c}^*
\end{align}
to estimate $\bbeta$?  The answer is yes. Here, $\tbeta_c$ is not only unbiased
but also has a smaller variance compared with $\tbeta$ in \eqref{eq:2}.

The unbiasedness of $\tbeta_c$ has been noticed in \cite{Yu2022}. We show it
here for completeness. Note that
\begin{align}
  \y_{c}^*
  =\y^*-\bar{y}\1_r
  =\alpha\1_r+\X^*\bbeta+\bve^*-(\alpha+\bar{\x}\tp\bbeta+\bar\varepsilon)\1_r
  =\X_{c}^*\bbeta+\bve^*-\bar\varepsilon\1_r,
\end{align}
where $\bar\varepsilon$ is the average of $\varepsilon_1, ..., \varepsilon_n$
and therefore $\bar\varepsilon\sim\Nor(0,n^{-1}\sigma^2)$. 
We then know that
\begin{align}
  \tbeta_c
   =(\X_{c}^{*\tsp} \X_{c}^*)^{-1}
    \X_{c}^{*\tsp}(\X_{c}^*\bbeta+\bve^*-\bar\varepsilon\1_r)
  & =\bbeta + (\X_{c}^{*\tsp} \X_{c}^*)^{-1}\X_{c}^{*\tsp}(\bve^*-\bar\varepsilon\1_r).
\end{align}
Thus, we have the unbiasedness $\Exp(\tbeta) = \bbeta$ from the above
representation.

The following proposition shows that $\tbeta_c$ has a smaller variance than
$\tbeta$ in the Loewner order.
\begin{proposition}\label{prop:1}
  Assume that $\Z^*$ is full rank. Let $\bar{\x}^*$ be the vector of
subsample covariate means, i.e., $\bar{\x}^*=r^{-1}\X^{*\tsp}\1_r$. The variances of $\tbeta_c$ and $\tbeta$ satisfy that
\begin{equation}\label{eq:16}
\Var(\tbeta\mid\X) - \Var(\tbeta_c\mid\X)
  =\sigma^2\bigg(\frac{r}{1-d}+\frac{r^2}{n}\bigg)
   (\X_{c}^{*\tsp}\X_{c}^*)^{-1}(\bar{\x}^*-\bar{\x})^{\otimes2}
    (\X_{c}^{*\tsp} \X_{c}^*)^{-1},
\end{equation}
where 
$d =
r(\bar{\x}^*-\bar{\x})\tp(\X_{c}^{*\tsp}\X_{c}^*)^{-1}(\bar{\x}^*-\bar{\x})\tp<1$.
\end{proposition}

\begin{remark}
  The smaller variance of $\tbeta_c$ indicates that even if the full data is not
  centered, it would be better to shift the subsample by $(\bar{\x},\bar{y})$
  and then fit a model without the intercept than fitting a model with an
  intercept directly.
\end{remark}

\begin{remark}
  The matrix on the right hand side of \eqref{eq:16} is of rank one,
  so we do not expect the difference between $\Var(\tbeta\mid\X)$ and
  $\Var(\tbeta_c\mid\X)$ to be large, especially when the dimension of
  $\bbeta$ is high. Nevertheless, we recommend $\tbeta_c$ because its
  calculation is as easy as that of $\tbeta$.
\end{remark}

If the intercept $\alpha$ is of interest, it can be estimated by using the
subsample means $\bar{\x}^*$ and $\bar{y}^*$. However, using the full data means
$\bar{\x}$ and $\bar{y}$ is usually significantly more efficient. This is also
observed in the numerical examples of
\cite{WangYangStufken2019}. Specifically, the estimator
\begin{equation}
    \talpha_c=\bar{y}-\bar{\x}\tp\tbeta_c
\end{equation}
is typically much more efficient than $\talpha$ defined in~\eqref{eq:2}. By direct calculations, we obtain that
\begin{equation*}
\Var(\talpha_c\mid\X)
  =\sigma^2\Big\{\onen + \bar{\x}\tp\Var(\tbeta_c\mid\X)\bar{\x}\Big\}
    \quad\text{and}\quad
  \Var(\talpha\mid\X)
  =\sigma^2\Big\{\oner+\bar{\x}^{*\tsp}\Var(\tbeta\mid\X)\bar{\x}^*\Big\}.
\end{equation*}
\cite{WangYangStufken2019} have shown that $\Var(\tbeta\mid\X)$ converges to
zero faster than $r^{-1}$ if the support of the covariate distribution is not
bounded. For this scenario, the dominating term in
$\Var(\talpha_c\mid\X)$ is often
$\bar{\x}\tp\Var(\tbeta_c\mid\X)\bar{\x}$ which converges to zero faster than
$\sigma^2r^{-1}$, while the dominating term in $\Var(\talpha\mid\X)$ is
$\sigma^2r^{-1}$.

\section{Nonuniform random subsampling with WLS}
\label{sec:weighted-ls}

A large class of subsample selection methods are through nonuniform random
sampling such as the leverage sampling
 and its variants \citep{PingMa2015-JMLR, yang2015explicit}, robust active
sampling \citep{nie2018minimax}, and optimal sampling
\citep{zhang2021optimal}. 
In this scenario the inverse
probability WLS approach is typically applied on the
subsample, and the subsample estimator is often proposed as an ``estimator'' of
the full data estimator.  Subsampling methods in this categorical may not
be as efficient as the deterministic selection methods in the previous section
if the assumed linear model is the data generating model,
but they require weaker assumptions on the correctness of the model.  For this type of approaches, the exact variance of the resulting
estimator may not be defined, so our discussions are on the asymptotic variance
which we use $\Var_a$ to denote. Properties of random subsampling estimators
are more complicated and we focus on the scenario that $r=o(n)$ so that the
contribution of the randomness from the full data to the asymptotic variance is
negligible.

Assume that a subsample of size $r$ is randomly selected according to nonuniform
probabilities $\pi_1, ..., \pi_n$, where $\pi_i$ is the probability that
the $i$-th observation is selected in each sample draw. For leveraging sampling,
$\pi_i$'s are proportional to statistical leverage scores; for robust active
sampling and optimal sampling, $\pi_i$'s are derived to minimize the asymptotic
mean squared error under mis-specified and correctly specified models,
respectively.  Here we abuse the notations and use $\X^*$,
$\y^*$, $\bve^*$, and $\Z^*$ again to denote subsample quantities. We need to
point out that \eqref{eq:3} does not hold for a randomly selected subsample.  

Let $\w=(w_1, ..., w_n)\tp$ be the vector of weights where $w_i$'s are
proportional to $\pi_i^{-1}$'s. To ease the discussion, we assume that $\|\w\|=1$. Let $\W$ be the corresponding $n\times n$ diagonal weighting matrix, i.e.,
$\w=\W\1_n$, and let $\w^*$ and $\W^*$ be the weighting vector and matrix, respectively, for the
selected subsample. The WLS estimator is
\begin{equation}\label{eq:15}
\ttheta_w = (\talpha_w,\tbeta_w\tp)\tp
  = (\Z^{*\tsp}\W^*\Z^*)^{-1}\Z^{*\tsp}\W^*\y^*.
\end{equation}

It has been shown that $\ttheta_w$ is asymptotically unbiased towards the full
data OLS $\htheta$, and its asymptotic variance has been derived in the
literature \cite[see][etc]{PingMa2015-JMLR, yu2020quasi, WangZouWang2022}. In our
notation, the asymptotic variance of $\ttheta_w$ given $\Dn$ is
\begin{align}
  \Var_a(\ttheta_w\mid\Dn)
  = \frac{C}{r}(\Z\tp\Z)^{-1}\Z\tp\W\E^2\Z(\Z\tp\Z)^{-1},
\end{align}
where $C=\sumn w_i^{-1}$ and $\E=\text{diag}(e_1, ..., e_n)$ with $e_i$'s being
the residuals from the full data OLS estimator. Note that $\Z\tp\W\E^2\Z=\sumn
w_ie_i^2\z_i\z_i\tp$, so if the weights $w_i$'s do not involve
$e_i$'s, then $\Z\tp\W\E^2\Z=\sigma^2\sumn
w_i\z_i\z_i\tp\{1+\op\}$ under reasonable conditions. Thus the asymptotic
variance can be written as
\begin{align}\label{eq:4}
  \Var_a(\ttheta_w\mid\Dn)
  = \frac{C\sigma^2}{r}(\Z\tp\Z)^{-1}\Z\tp\W\Z(\Z\tp\Z)^{-1},
\end{align}
from which %
we obtain that
\begin{align}\label{eq:5}
  \Var_a(\tbeta_w\mid\Dn)
  = \frac{C\sigma^2}{r}
  (\X_{c}\tp\X_{c})^{-1}\X_{c}\tp\W\X_{c} (\X_{c}\tp\X_{c})^{-1}.
\end{align}
If the centered data is sampled and used to construct an estimator of $\bbeta$
directly
\begin{align}\label{eq:7}
  \tbeta_{w,uc} = (\X_c^{*\tsp}\W^*\X_c^*)^{-1}\X_c^{*\tsp}\W^*\y_c^*,
\end{align}
then since the full data means $\bar\x$ and $\bar{y}$ are nonrandom functions of
the full data $\Dn$, existing results \cite[e.g.,][]{ai2020optimal, yu2020quasi,
  WangZouWang2022}
are applicable to $\tbeta_{w,uc}$. This tells us that $\tbeta_{w,uc}$ is
asymptotically unbiased towards $\hbeta$ and its asymptotic variance is the same
as that of $\tbeta_w$ shown in~\eqref{eq:5}. Thus for noninformative
random subsampling with WLS, if the original full data is centered, we can
ignore the intercept as well. %

Interestingly, if we use weighted means of the full data to relocate the
subsample, we have an improved estimator of $\bbeta$. Denote $\bar{y}_w=\w\tp\y$
and
$\bar{\x}_w=\X\tp\w$ as the weighted mean response and the weighed mean covariate
vector, respectively. Let $\y_{wc}=\y-\bar{y}_w\1_n$ and
$\X_{wc}=\X-\1_n\bar{\x}_w\tp$ be the centered response vector and design matrix
using the weighted means, respectively, and let $\y_{wc}^*$ and $\X_{wc}^*$ be
the corresponding selected subsample quantities. A better subsample estimator
for $\bbeta$ is
\begin{align}
  \tbeta_{wc} = (\X_{wc}^{*\tsp}\W^*\X_{wc}^*)^{-1}\X_{wc}^{*\tsp}\W^*\y_{wc}^*.
\end{align}
Again, since $\bar{y}_w$ and $\bar{\x}_w$ are nonrandom functions of the full
data $\Dn$, existing results show that $\tbeta_{wc}$ is asymptotically unbiased
with asymptotic variance
\begin{align}\label{eq:6}
  \Var_a(\tbeta_{wc}\mid\Dn)
  &= \frac{C\sigma^2}{r}
   (\X_{wc}\tp\X_{wc})^{-1}\X_{wc}\tp\W\X_{wc}
    (\X_{wc}\tp\X_{wc})^{-1}.
\end{align}
The following result shows that $\tbeta_{wc}$ has a smaller asymptotic variance
than $\tbeta_w$.
\begin{proposition}\label{prop:2}
The asymptotic variances in \eqref{eq:5} and \eqref{eq:6} satisfy that
$\Var_a(\tbeta_{wc}\mid\Dn)\le\Var_a(\tbeta_w\mid\Dn)$, and the equality holds
if $\bar{\x}_{wc}=\bar{\x}$. 
\end{proposition}

\begin{remark}
  The above result relies on the asymptotic representation in \eqref{eq:4} which
  requires that the weights do not involve the residuals. Since the weights are
  inversely proportional to the sampling probabilities, this means that the sampling
  probabilities are noninformative, i.e., they do not
  depend on the responses. For informative subsampling such as the A- or L-
  optimal subsampling \citep{ai2020optimal,WangZouWang2022}, there is no definite ordering between
  $\Var_a(\tbeta_{wc}\mid\Dn)$ and $\Var_a(\tbeta_w\mid\Dn)$.
\end{remark}

Similar to the case of deterministic selection, if the intercept is of interest,
it can be estimated by
\begin{align}
  \talpha_{wc} = \bar{y}_w - \bar{\x}_w\tp\tbeta_{wc}
  \quad\text{or}\quad
  \talpha_{w,uc} = \bar{y} - \bar{\x}\tp\tbeta_{wc}.
\end{align}
Note that the $\talpha_w$ defined in~\eqref{eq:15} satisfies
$\talpha_w=\bar{y}_w^* - \bar{\x}_w^*\tbeta_w$. Both $\bar{y}_w^*$ and
$\bar{\x}_w^*$ are random given $\Dn$, and thus they both contribute to the
variation of $\talpha_w$ in approximating $\halpha$. For $\talpha_{wc}$ (or
$\talpha_{w,uc}$), neither $\bar{y}_w$ nor $\bar{\x}_w$ (or $\bar{y}$ nor
$\bar{\x}$) is random given $\Dn$, so the only source of variation in
approximating $\halpha$ is $\tbeta_{wc}$. Thus $\talpha_{wc}$ and
$\talpha_{w,uc}$ are often significantly more efficient than $\talpha_w$.

\section{Numerical comparisons}
\label{sec:numer-comp}

We provide some numerical simulations that compare the performance of
the estimators discussed in previous sections. We generated data from
model~\eqref{eq:1} with $n=10^5$, $\alpha=1$, $\bbeta=\1_{19}$, and
$\bve\sim\Nor(\0,\sigma^2\I_n)$ with $\sigma^2=9$. To generate rows of $\X$, we
considered the following three distributions. Case 1: multivariate normal
distribution $\Nor(\0,\bSigma)$, Case 2: multivariate log normal distribution
$\exp\{\Nor(\0,\bSigma)\}$, and Case 3: multivariate $t$ distribution with
degrees of freedom five $\mathbb{T}(\0,\bSigma,5)$. Here the $(i,j)$-th element of
$\bSigma$ is $0.5^{|i-j|}$ in all cases.
 These cases were used because the skewness and tail heaviness of the
covariate distribution have a significant impact on the performance of different
subsampling methods. Case 1 is a light-tailed distribution and Case 3 is a
heavy-tailed distribution. Existing asymptotic results on the subsampling
methods mentioned in Section~\ref{sec:weighted-ls} require the forth moment of
the covariate distribution to be finite, and Case 3 gives the heaviest tail
within available asymptotic frameworks. Cases 1 and 3 are symmetric
distributions while Case 2 is an asymmetric distribution.  
 We implemented three subsampling
methods: uniform sampling, IBOSS \citep{WangYangStufken2019}, and leverage
sampling \citep{PingMa2015-JMLR}. We run the simulation for 1,000 times to
calculate the empirical mean squared errors (MSE) reported in Table~\ref{tab:1}.

We see that for the slope parameter, although not very significant, an estimator
based on centered full data (WOI, un-centered subsample) has a smaller MSE than the
counterpart based on un-centered full data (WI, centered subsample). For the
intercept, the estimator based on the full data means (WOI) is better than the
counterpart based on the subsample only (WI), and the improvement is quite
significant.  Furthermore, by comparing results for Case 1 and Case 3, we
see that the improvement is more significant if the covariate distribution has a
heavier tail. One reason is that the variation of the slope estimator also
contributes to the MSE of the intercept estimator and the slope estimator has a
smaller variance if the covariate distribution has a heavier tail. The covariate distribution
is asymmetric in Case 2, so the leverage sampling gives higher preferences for
data points in one tail of the full data and the means of the selected subsample
do not perform well. Using the full data means to replace the subsample means
resulted in the most significant improvement for estimating the intercept in this
case. 
 
\begin{table}[htbp]
  \caption{Empirical MSEs of subsample estimators for the intercept and slope.$^{*}$}
\label{tab:1}
\centering
\begin{tabular}{ccrrcrrcrr}
  \hline
 & & \multicolumn{2}{c}{Uniform} && \multicolumn{2}{c}{IBOSS} && \multicolumn{2}{c}{Leverage} \\ \cline{3-10}
 && \multicolumn{1}{c}{WI} & \multicolumn{1}{c}{WOI} &&
 \multicolumn{1}{c}{WI} & \multicolumn{1}{c}{WOI} &&
 \multicolumn{1}{c}{WI} & \multicolumn{1}{c}{WOI} \\ \hline
Case 1 & $\alpha$ & 88.422 & 76.342 && 27.094 & 26.608 && 83.463  & 73.260 \\ 
       & $\bbeta$ & 18.745 & 18.482 && 12.874 & 12.740 && 17.834  & 17.641 \\ 
Case 2 & $\alpha$ & 81.065 & 79.155 && 12.693 & 12.579 && 152.350 & 22.085 \\ 
       & $\bbeta$ & 0.683  & 0.668  && 0.064  & 0.062  && 0.400   & 0.396  \\ 
Case 3 & $\alpha$ & 54.336 & 45.025 && 1.505  & 0.655  && 67.008  & 15.929 \\ 
       & $\bbeta$ & 13.418 & 13.259 && 0.564  & 0.558  && 10.520  & 10.500 \\ \hline
  \multicolumn{10}{l}{\footnotesize $^{*}$ ``WI'': $\alpha$ and $\bbeta$ are
  estimated from a model with an intercept; }\\[-1mm]
  \multicolumn{10}{l}{\footnotesize``WOI'': $\bbeta$ is estimated from a model
  without an intercept, and}\\[-2mm]
  \multicolumn{10}{l}{\footnotesize\hspace{1.2cm} $\alpha$ is estimated using
  the full data means.}\\
\end{tabular}
\end{table}

\section{Summary}
For a subsample selected from centered full data, although the subsample is
un-centered, it is better to fit a model without an intercept to estimate the
slope parameter if the subsampling rule does not depend on the response
variable. If the full data is un-centered, it would be better to shift the
location of the data by the full data (weighted) means for the OLS (WLS) and
then fit a model without an intercept.

\section*{Acknowledgement}
The author thanks the associate editor and referee for
their comments which helped improve of this article.
  This work is supported by NSF grant CCF 2105571.
  
\appendix
\section{Technical details}
\label{sec:technical-details}
\begin{proof}[\bf Proof of Proposition~\ref{prop:1}]
  Let $\S$ be the $r\times n$ selection matrix consisting of zeros and ones that
  maps the full data to the subsample, i.e., $\y^*=\S\y$ and $\X^*=\S\X$. We know
  that 
\begin{align}
  \X_{c}^*&=\X^*-\1_r\bar{\x}\tp =\X^*-n^{-1}\1_r\1_n\tp\X =\S(\I_n-\J_n)\X,\\
  \y_{c}^*&=\y^*-\bar{y}\1_r = \S\y-n^{-1}\1_r\1_n\tp\y =\S(\I_n-\J_n)\y.
\end{align}
Thus, 
\begin{align}
  \Var(\y_{c}^*\mid\X) = \Var\{\S(\I_n-\J_n)\bve)\} = \sigma^2(\I_r-rn^{-1}\J_r),
\end{align}
and therefore
\begin{align}
  \Var(\tbeta_c\mid\X)
  & = (\X_{c}^{*\tsp} \X_{c}^*)^{-1}\X_{c}^{*\tsp} \Var(\y^*\mid\X)
    \X_{c}^*(\X_{c}^{*\tsp} \X_{c}^*)^{-1}\\
  & = \sigma^2(\X_{c}^{*\tsp} \X_{c}^*)^{-1}\X_{c}^{*\tsp} (\I_r-rn^{-1}\J_r)
    \X_{c}^*(\X_{c}^{*\tsp} \X_{c}^*)^{-1}\\
  & = \sigma^2(\X_{c}^{*\tsp} \X_{c}^*)^{-1}
    - \sigma^2rn^{-1}(\X_{c}^{*\tsp} \X_{c}^*)^{-1}\X_{c}^{*\tsp} \J_r
    \X_{c}^*(\X_{c}^{*\tsp} \X_{c}^*)^{-1}.\label{eq:9}
\end{align}
Let $\X_{cr}^*$ be the subsample design matrix centered by the subsample
means, i.e., $\X_{cr}^*=\X^*-\1_r\bar{\x}^{*\tsp}$. From the facts that $\1_r\tp\X_{cr}^*=\0\tp$ and
\begin{align}\label{eq:11}
  \X_c^* = \X^*-\1_r\bar{\x}\tp = \X_{cr}^*-\1_r(\bar{\x}-\bar{\x}^*)\tp,
\end{align}
we know
\begin{align}
    \X_{c}^{*\tsp} \J_r\X_{c}^*
  = r^{-1}\X_{c}^{*\tsp} \1_r\1_r\tp\X_{c}^*
  = r(\bar{\x}-\bar{\x}^*)^{\otimes2}.\label{eq:10}
\end{align}
Thus \eqref{eq:9} and \eqref{eq:10} give
\begin{align}\label{eq:12}
  \Var(\tbeta_c\mid\X)
   = \sigma^2(\X_{c}^{*\tsp} \X_{c}^*)^{-1}
    - \sigma^2r^2n^{-1}(\X_{c}^{*\tsp} \X_{c}^*)^{-1}(\bar{\x}-\bar{\x}^*)^{\otimes2}(\X_{c}^{*\tsp} \X_{c}^*)^{-1}.
\end{align}

From \eqref{eq:2}, the variance of $\tbeta$ is
\begin{align}\label{eq:13}
    \Var(\tbeta\mid\X) =\sigma^2(\X_{cr}^{*\tsp} \X_{cr}^*)^{-1}.
\end{align}
Note that~\eqref{eq:11}
implies
\begin{align}\label{eq:8}
  \X_{cr}^{*\tsp}\X_{cr}^*
  &= \X_{c}^{*\tsp} \X_{c}^* - r\{\bar{\x}^*-\bar{\x}\}^{\otimes2}.
\end{align}
Thus we obtain 
\begin{align}\label{eq:14}
  (\X_{cr}^{*\tsp}\X_{cr}^*)^{-1}
  = (\X_{c}^{*\tsp} \X_{c}^*)^{-1}
    +\frac{r(\X_{c}^{*\tsp}\X_{c}^*)^{-1}(\bar{\x}^*-\bar{\x})^{\otimes2}
    (\X_{c}^{*\tsp} \X_{c}^*)^{-1}}
    {1-d},
\end{align}
where $d=r(\bar{\x}^*-\bar{\x})\tp(\X_{c}^{*\tsp}\X_{c}^*)^{-1}
(\bar{\x}^*-\bar{\x})\tp$.  Here $1-d$ must be positive because \eqref{eq:8}
implies that $(\X_{cr}^{*\tsp}\X_{cr}^*)^{-1} \ge (\X_{c}^{*\tsp}
\X_{c}^*)^{-1}$ and they are both positive-definite.

Combining \eqref{eq:12}, \eqref{eq:13}, and \eqref{eq:14} finishes the proof.
  
\end{proof}

\begin{proof}[\bf Proof of Proposition~\ref{prop:2}]
For positive definite matrices $\A_1$, $\A_2$, $\B_1$, and
$\B_2$, if $\A_1\le\A_2$ and $\B_1\ge\B_2$, then
\begin{align*}
  \B_2^{1/2}\A_1^{-1}\B_2^{1/2} \ge \B_2^{1/2}\A_2^{-1}\B_2^{1/2}
  \Rightarrow
  & (\B_2^{1/2}\A_1^{-1}\B_2^{1/2})^2 \ge (\B_2^{1/2}\A_2^{-1}\B_2^{1/2})^2\\
  \Rightarrow
  & \A_1^{-1}\B_2\A_1^{-1} \ge \A_2^{-1}\B_2\A_2^{-1}, %
\end{align*}
so $\A_1^{-1}\B_1\A_1^{-1} \ge \A_1^{-1}\B_2\A_1^{-1} \ge
\A_2^{-1}\B_2\A_2^{-1}$. 
Thus we only need to prove that
\begin{align}
  \X_{c}\tp\X_{c}   &\le \X_{wc}\tp\X_{wc}, \\
  \X_{c}\tp\W\X_{c} &\ge \X_{wc}\tp\W\X_{wc},
\end{align}
and the equality in both hold if $\bar{\x}_{wc}=\bar{\x}$. The proof finishes
from the fact that
\begin{align}
  &\X_{c}\tp\W\X_{c} - \X_{wc}\tp\W\X_{wc}
    = \X\tp(\w-n^{-1}\1_n)^{\otimes2}\X
    = (\bar{\x}_{wc}-\bar{\x})^{\otimes2} \ge\0,\\
  &\X_{wc}\tp\X_{wc} - \X_{c}\tp\X_{c}
    = n\X\tp(\w-n^{-1}\1_n)^{\otimes2}\X
    = n(\bar{\x}_{wc}-\bar{\x})^{\otimes2} \ge\0,
\end{align}
which can be verified by inserting 
$\X_{c}=(\I_n-n^{-1}\1_n\1_n\tp)\X$ and $\X_{wc}=(\I_n-\1_n\w\tp)\X$.
\end{proof}

\bibliographystyle{agsm}
\bibliography{ref}

\end{document}